# Has our brain grown too big to think effectively?

**Konrad R. Fialkowski**

University of Warsaw, Poland.

Mailing address: An den Langen Luessen 9/1/3; 1190 Wien, Austria
Fax: 431 3288689
e-mail:       fialkows@aol.com

**Abstract.**

A variant of *microcephalin, MCPH1* gene, was introgressed about **37,000** years ago into *Homo sapiens* genetic pool from an archaic (*Homo erectus*) lineage and rose to exceptionally high frequency of around **70 percent worldwide** today. It is involved in regulating neuroblast proliferation and its changes alter the rate of division and/or differentiation of neuroblasts during the neurogenic phase of embriogenesis, which could alter the size and structure of the resulting brain.

At the time of introgression, images had already been painted on the walls of caves and speech has been in use for over 100,000 years, as had been abstract thinking. Like today, reasoning and thinking were the primary faculties of individuals. *Homo erectus* either did not possess those faculties or was markedly inferior to *Homo sapiens* in them. Its brain was smaller and the cortex was apparently less convoluted. Thus, introgressed *microcephalin* allele directed neurogenesis evolutionary back to less complicated brain structure typical for our evolutionary forefathers, slightly decreasing the level of complexity already achieved by *Homo sapiens* 37,000 years ago. Despite that, it proliferated at a rapid pace.

It yields a supposition: 37,000 years ago the brains of *Homo sapiens* were too big and too complicated for the kind of thinking needed for the highest fitness of individuals. Since adaptation cannot by definition surpass selection requirements, **the volume and complication of the human brain did not originate under selective pressure to improve effective thinking** and they cannot **be explained in terms of such selection.**

A proposal to solve this quandary is presented, claiming **that *Homo sapiens* originated just by chance**. Endurance running led to the emergence of *Homo sapiens*. The human mind and larynx used for speech are side-effects of more than a million years of endurance running by pre-human hunters.

**Introgression of the *Microcephalin* gene.**

Recent research findings and ever advancing knowledge of the genome structure associated with the human brain and its evolution has yielded compelling evidence that genes exert a major influence on the volume of the human brain (Tang; 2006) as well as on the size and morphology of its functional domains (Hill et al., 2005).

Among the genes identified as the regulators of brain size, the gene known as *microcephalin* – MCPH1 – is the subject of further consideration in this paper. The influence of *microcephalin* (together with *ASPM* gene) on brain size and structure is stressed by Gilbert et al. (2005: 6):

"If ASPM and MCPH1 are involved in regulating neuroblast proliferation, then evolutionary changes in these genes might alter the rate of division and/or differentiation of neuroblasts during the neurogenic phase of embriogenesis, which in turn could **alter the size and structure of the resulting brain**". (Emphasis added).

This paper focuses on the *microcephalin* gene because a genetic variant of *microcephalin* was introgressed into the modern human genome approximately **37,000** years ago (Evans et al., 2006) and proliferated rapidly in the modern human gene pool. This allele originated from a lineage separated from modern humans **about 1.1 million** years ago. The rapid proliferation of the allele seems to be a crucial indicator of selection forces transforming the hominid brain into the contemporary human brain.

Evans et al., (2006) announced that the adaptive allele of the brain size gene *microcephalin* introgressed into the *Homo Sapiens* population from an archaic *Homo* lineage, increasing its frequency from a single copy (aproximately 37,000 years ago) to an exceptionally high **frequency of about 70% worldwide today.**

It should be added that the exchange of genes between species long after they branched off from a common ancestor is standard for sympatric speciation. Gene exchange between chimpanzees and early hominids long after divergence was found by Patterson et al. (2006).

This feature is anything but unexpected. The theory of sympatric speciation (Fialkowski, 1992) indicates that in the process, lineages initially diverge, then, later, exchange genes, before separating permanently. This pattern was also confirmed in another case of diverging species (Fialkowski, 1994).

Without excluding *Homo erectus*, Evans et al. indicate Neanderthal as a source of the introgressed allele. This issue is essential for further considerations presented here. Another case of introgression to the human genetic pool was indicated by Garrigan et al. (2005). They refer to the introgression of *RRM2P4* that originated from the Asiatic branch of *Homo erectus*.

The most convincing indication that *Homo erectus* was the source of the introgressed *microcephalin* allele stems from Reed et al. (2004). They found that modern human head lice, *Pediculus humanus*, comprised two ancient lineages that diverged about **1.18 million** years ago. One lineage has a worldwide distribution and apparently underwent (about 100,000 years ago) a population bottleneck together with its modern *Homo sapiens* host. The other lineage that apparently changed the host from an extinct *Homo* to *Homo sapiens* is solely found in Americas. Its absence in Europe rather excludes Neanderthal as its host. It is speculated that Asian migrants during the late Stone Age have transported the latter lineage to America.

This would indicate that physical contacts probably occurred between *Homo erectus* and *Homo sapiens,* with eastern Asia being the most likely geographical region for such contacts some 50,000 - 25,000 years ago. It coincides with the period when *microcephaline* allele was introgressed.

Moreover, the time when the lice lineage split (approximately 1.18 million years ago) closely coincided with the time of the split for the *microcephaline* alleles (about 1.1 million years ago). These values, independently, indicate the same time for the occurrence of speciation in the *Homo* lineage. Together with the research findings of Garrigan et al. (2005) quoted above pertaining to the other introgression stemming from *Homo erectus*, this strongly suggests that *Homo erectus* was indeed the source of introgression *microcephaline* allele into the human genetic pool.

**The paradox of the *microcephalin* introgression.**

The paradox arises not from introgression of the *microcephalin* allele itself, but from its proliferation in the modern human population. The range of proliferation implies that this particular allele had led to a **higher fitness** in individuals who were **shifting back to that allele of their evolutionary forefathers**.

At the time of introgession (about 37,000 years ago), images had already been painted on the walls of caves and burial ceremonies performed. Speech had been in use from more than 100,000 years (Liberman, 1991) to the accompaniment of abstract thinking. Generally, similar to today, reasoning and thinking were the primary faculties of individuals.

*Homo erectus* did not possess those faculties, or at least was markedly inferior in all of them. Its brain was smaller and cortex apparently was less convoluted. As *microcephalin*, quoting Evans et al, (2006: pp. 18178):

"..likely plays an essential role in promoting the proliferation of neural progenitor cells during neurogenesis"**,**

the neurogenesis stemming from *microcephalin* allele 'possessed' by *Homo erectus* rendered a simpler, less complicated brain, directing neurogenesis evolutionary back towards a structure that was more similar to the structure of brains of our evolutionary forefathers than those already achieved by *Homo sapiens* 37,000 years ago. Despite all that, proliferation was broad ranged. That is paradoxical.

If this reasoning is correct, only one conclusion can be drawn:

**37,000 years ago, the brains of *Homo sapiens* were too big and too complicated for the kind of thinking that could have ensured the highest fitness of individuals.**

Ignoring everything we accept as obvious, one justified question, never posed before, can be asked:

Why, generally speaking, did a technological/organizational progress begin so late, some 15,000 years ago and not earlier, even though by that time modern *Homo sapiens* displaying speech and other faculties had already been in existence for at least 100,000 years?

Perhaps, the introgressed *microcephalin* allele brought about the change.

It is possible to estimate roughly the percentage of the population with introgressed *microcephalin* allele at the onset of technological progress, yet the calculation was never made. However, given that the proliferation of the allele is exponential in character, a good guess might be in the order of 10 – 15%, which is close to the tipping-point fraction of a population needed for a spread of phenomena in networks (Xie et al. 2011).

There are also some indications (Bielicki, 2001: 31) that, in statistical term, the volume of the human brain shrunk slightly over the past 30,000 years.

Could it be so that those individuals from the time prior to the introgression, who painted on the rocks and could possibly be termed "a population of artists" differed in essence from those with the introgressed allele - more technologically oriented people – "the techno population". The strength of the selection clearly indicates the essential difference between the two groups. Whatever the difference was, it has been rendering essentially different fitness, i.e., essentially better survival and/or fertility.

In a sense, the hypothesis presented is falsifiable. If correct, it implies a positive correlation between a **less convoluted cortex in youth** and more effective pragmatic thinking.

The introgression discussed occurred at a time when use of the brain in a human manner was already the primary and most important means of maintaining high fitness of individuals.

Proliferation of the introgressed allele 'taken' from the mentally inferior hominid with its smaller and less complicated brain indicates that a decrease in the complexity of the brain rendered higher fitness. The broad proliferation of the introgressed allele indicates substantial gain in fitness brought about by the allele.

It thus follows that prior to introgression, the human brain was too big and too complicated to think most effectively or at least the smaller brain developed after the introgressed allele was able to think equally effectively.

This means that the **volume and complexity of the human brain did not originate under selective pressure to improve effective thinking**, because adaptation cannot by definition surpass selection requirements. Given the range to which the introgressed allele proliferated, it is evident that the shift to the less complicated brain, more similar to the brains of our evolutionary forefathers, brought about better individual fitness.

That brings back the quandary that François Jacob (1977: 1166) articulated in the lecture he delivered at the University of California:

"Although our brain represents the main adaptive feature of our species, what it is adapted to is not clear at all."

This quandary remains unresolved. An attempt to answer it has been undertaken (Fialkowski 1978, 1986, 1994) and the final results were recently published in a book *HOMO perchance SAPIENS* (Fialkowski & Bielicki 2008).

**Humans by chance.**

It is claimed that *Homo sapiens* originated just by chance. The scenario for his origin was highly improbable, yet it happened none the less! No ecological imperative determined the advent of *Homo sapiens*.

The question **why** it happened cannot by definition be answered in evolutionary terms.

The question **how** it happened is not easy to answer.

The minds of individuals belonging to our species are unique. We perceive reality in the same way as other species do, yet we also (and in fact, mainly) perceive reality through words (including among those words the representation of time: something that can hardly be represented in image form without words), while chimpanzees, our closest evolutionary relatives, take many years learning how to crack nuts using stones: an achievement that comes close to the limits of their intellectual capabilities.

The faculties of our species, however, differ from those of all other species of mammals in one other specific way that passes almost unheeded in modern society. Whereas our minds are intensely exploited, the second specifically human faculty now remains practically unused: comparable in its obsolescence to a tinderbox.

The faculty in question is our manner of running, more specifically the bipedal jump: a strange and unique mechanism essentially different from the running patterns of all other mammalian species. Humans can run uninterruptedly for hours (viz. marathon running). Despite its low speed, this persistence or endurance running (when imposed on animals) exceeds the stamina and capabilities of dogs, kangaroos and even pronghorn antelopes, one of the fastest mammals. Under certain thermal conditions, it also exceeds the capacity of a zebra. All these animals ultimately collapse when pursued and forced by man to run for long enough periods, whereupon they fall easy prey to human hunters.

Despite it being only a marginal trait of current human behaviour, **endurance running led to the emergence of *Homo sapiens*. The human mind and larynx used for speech are the evolutionary 'reward' for more than a million years of endurance running by pre-human hunters.**

The substantiation of this approach, however, calls for a step-by-step scenario that explains in terms of evolutionary biology each consecutive step of the transformation process leading from *Australophitecine* - the first ape-like hominid that lived several million years ago - to the human being. Each transformational step in that scenario has to be justified in keeping with the rules of the evolutionary process and show the manner in which those rules were applied.

Evolution is a unique process, in the course of which species are shaped by environmentally determined forces of selection. Just as a block of stone potentially contains all sculptures of the world, so too do all existing species contain their potential adjustment capabilities. When the form carved from stone exceeds the durability or endurance of the material, the sculpture shatters. When environmental determinants push a species beyond its adjustment capabilities, the species becomes extinct. Along the borders of 'material endurance' the strangest shapes and forms are to be found- forms that come close to destruction, yet do not reach it. On at least one occasion, the emerging human species came extremely close to extinction.

When its forest habitat began to disappear, the bipedal ape found itself on the savannah. With limited chances of survival, it was forced to join the predator's guild, even though it lacked fangs and claws: the attributes of the guild. Moreover, to its disadvantage, the bipedal ape found that all the traditional hunting niches were already occupied. The predators with fangs and claws had already staked their claims to morning, evening and night hunts. The bipedal ape had to take the only option left: the diurnal hunt, when the tropical sun was at its hottest and the well-established hunters were taking a nap.

The sole trump card that the bipedal ape could play was its **specific long-distance endurance running capability** acquired through its bipedality. That specific form of running became its 'antlers' of effectiveness comparable to fangs and claws. Even today some tribes still hunt in that manner: trotting relentlessly after their prey, allowing it no rest until it is completely exhausted and the coup de grace is simple. Under these circumstances, a weapon is not necessarily required for the kill (even hands suffice) and the game won by a single hunter is greater than the booty of a group hunt. The bipedal ape managed to survive. Hominid, our pre-ancestor became a dangerous predator.

Endurance hunting also led to an increase in volume of the hominid brain; its larger and more interconnected structure was an improvement in terms of its resistance to the overheating caused by endurance running. The restructured brain acted as a **'reliability thermal buffer'**. The changed brain structure was less susceptible to overheating. Moreover, **quite by chance,** a side-effect was its greater adjustment to improved thinking.

In evolutionary biology such a case is termed pre-adaptation. The brain's thinking capabilities that emerged as a side effect of the improved **'reliability thermal buffer'** were eventually suitable for speech control and abstract thinking.

At the time this hypothesis was first published (Fialkowski 1978), the claim that the human brain was pre-adapted to speech and abstract thinking (i.e. the human brain did not originally evolve as adaptation to speech and abstract thinking, but later proved appropriate for both) was a conceptual *novum*: a proposal for a new paradigm for the origins of humankind. It contradicted the broadly accepted belief that the human brain originated as a response to requirements for better thinking.

According to the proposal (Fialkowski, 1978, 1986), hominid brain emerged as a result of pre-adaptation (similarly to lungs that emerged as an unexpected utilisation of the air-bladder in fish when they moved on to land). In pre-adaptation, a structure emerging as a result of a selection pressure is, by chance, appropriate for a new function that differs from the one which originated the selection pressure. Apart from adaptation, it is the second possibility offered by the Darwinian Theory. According to Mayr (1970: 423) a structure is pre-adapted, if it can assume a new function without interfering with the original function.

The hypothesis (Fialkowski, 1978, 1986) claims that:

(i) heat generated in hominid bodies during persistent hunting/running (Krantz 1968; Bortz 1985) was transported from the muscles via the blood stream to the brain, damaging neurones at random, impairing brain functions and decreasing hunting success. Effective blood cooling systems as in other mammals (Baker 1972, 1979; Baker & Chapman 1977) were not developed.

(ii) In terms of hunting success, the number of malfunctioning neurons in the brain tissue was irrelevant as long as the brain continued to function properly as a whole. Any variations in the brain structure, which increased the capability of the brain to maintain its function as a whole, despite some malfunctioning neurons, were strongly positively selected.

**(iii)** A reliability principle (von Neumann, 1963 [1952]) states that in order to increase the reliability of information structure composed of malfunctioning elements, both the number of elements and the number of connections between the elements must be increased. It is a general law applicable to all interconnected systems composed of discrete elements. **The reliability hypothesis claims that this principle found by von Neumann constituted a pattern for brain adaptation in hominids.**

Presuming that evolution "applied" von Neumann's principle to the pre-human brain, as a result of this adaptation, the adapted human brain should (Fialkowski 1990b):
    (i)    have an increased number of neurons;
    (ii)   the neurons should be more interconnected; and
    (iii)  the brain should be more resistant to heat stress.

Both features, (i) and (ii), deduced from von Neuman's theory are specific to reliability adaptation and can be found in the human brain: interconnectivity is greater than in the brain of great apes, and phylogenic growth of the brain volume is a manifestation of certain increase (1.25 times; Holloway 1966) in the number of neurons, which are less densely packed (Shariff 1953). The third feature, (iii), is also specific to the human brain. The human brain is clearly more resistant to heat stress than that of animals. As Brinnel et al. (1987: 209) put it (emphasis added):

"…in a view of the high levels of body temperature which have been recorded in runners (about 42° C) or in heat stroke patients (46.55° C), either there is a very appreciable extent of selective brain cooling or **the brain is much less temperature-sensitive than indicated by animal experiments**."

A heat-resistant brain enables primitive tribes, **even today**, to go persistence hunting. By trotting relentlessly and uninterruptedly after their prey, contemporary primitive hunters bring pronghorns (one of the fastest mammals) to the verge of collapse, as well as kangaroos and (when the day is hot) even zebras. Moreover, it is no coincidence that it is simply the brain's level of **resistance** to extreme blood temperatures that determines the **medallists among the best marathon runners**. The runner capable of sustaining the **highest temperature** wins the gold medal; the places thereafter are determined by the sequence of drops in the runners' blood temperatures.( Pugh et al. 1967, Hamilton 1973).

Heat stress as a primary selection factor (Fialkowski, 1978, 1986) fits this pattern well. The main source of heat, however, is a by-product of physical activity rather than sole exposition to sun radiation in a hot environment. As Bortz (1985: 148) stated it is "...the heat generated by exercise which is the discriminating burden". Thus, persistence hunting fulfils the requirements for the behaviour under drastic selection conditions that is required for the rapidly progressing adaptation discussed.

**Persistence hunting.**

Chasing prey (persistence hunting: Krantz 1968; Watanabe 1971; Carrier 1984; Bortz 1985; also quoted after Carrier: Schapera (1930); Bennet & Zingg 1935; Lowie 1924; Foster 1830; Solas 1924; McCarthy 1957) was a prevailing method of hunting in early hominids and despite the adoption of projectile weapons and other technologies by modern humans it is still in use today. The motives behind maintaining persistence hunting in contemporary human populations were presented by Bortz (1985:147):

"When prey density is low, individual hunting is wisest (Lamprecht, 1978). To obtain highest return per amount of time and energy expended in searching for a mobile resource, the best strategy would seem to be cover as much area as possible per person (Hayden, 1981). (...) chase myopathy renders any animal incapable of further retreat or defence so that individual hunting may have been very effective, indeed it could have been the predominant behaviour."

Hayden wrote, "Groups will hunt as individuals when they can and communally when they have to" (Hayden, 1981)".

Quoting Carrier (1984: 483): "Hunters of a number of **different cultures** [emphasis added] are known to run down prey by dogged pursuit often lasting one or two days (Kranz 1968, Watanabe 1971). Bushmen are reported to run down duiker, steenbok, and gemsbok during the rainy season and wildebeest and zebra during the hot dry season (Schapera 1930) [it is possible only in the hot season, as otherwise overheating of zebra and the chase myopathy that follows would not to be evoked – comments added]. Tarahumara Indians chase deer through the mountains of northern Mexico until the animals collapse from exhaustion and then throttle them by hand (Bennet and Zingg 1935, Pennington 1963). Paiutes and Navajo of the American Southwest are reported to have hunted pronghorn antelope (one of the fastest of all mammals) with this same technique (Lowie 1924, Foster 1830, cited by Lopez 1981: 111). Furthermore, the aborigines of north-western Australia are known to hunt kangaroo successfully in this way (Solas 1924, McCarthy 1957). In these examples the stamina of prey animals generally considered to be specialised for running and presumably putting forth their best effort proves to be less than that of man."

Also Lieberman (1991: 153), unintentionally, offers an example of apparent multi-culture presence of the persistent hunting that yielded the genetic fixation of the behaviour. To quote him, "The difference between chimpanzee and human moral sense can be seen most clearly in the hunt. Chimpanzees, like many humans, hunt other animals to acquire meat. [...] However, whereas human hunters kill their prey before they start to eat its flesh, chimpanzees do not seem to care whether the victim is dead or not." If chimpanzees were persistence hunters they would kill as well, because the chase myopathy is transient and lasts not too long.

It is worth to note that unlike other predators that usually select old, sick or young individuals as their prey, *Homo erectus* was not a selector. **Homo erectus was a killer**, for whom the physical condition of his pray was immaterial. By doggedly pursuing his prey throughout the hunt (chase myopathy) *Homo erectus* could catch and kill any animal, even the healthiest and the fittest.

**Redundancy and the exponential growth of the brain.**

In what way did this mode of hunting change the brain of the Homo? The brain is a complex system of interconnected **discrete** elements: neurons. Contrary to other organs (e.g. muscles), not all neurons are activated when an area of the brain is stimulated. Despite stimulation of the brain area, some of them do not transmit any signals. Moreover, stimulation generates a pattern of activation, not all the details of which are repeated in a subsequent identical stimulation. As a response to a stimulus (e.g. a flash of light), it can only be stated in statistical terms that most of the neurons in the stimulated area are activated. More specifically, however, at the level of the individual neurons in the stimulated area, the same neuron is not necessarily always stimulated - despite it's being always a part of this area.

It yields a conclusion that in a thought out experiment, selecting randomly a single neuron from the stimulated area, it is impossible to foresee, whether that neuron will in fact be stimulated in response to the stimuli. That, however, does not change the response of the brain as a whole to those stimuli. The flash is always recognized as a flash.

What would happen, were the neuron to disappear? Recognition of the flash would remain unchanged. The presence or absence of single neurons does not change the brain's reaction. In fact, every second of our life, a number of neurons die: an event we fail to register (at least for a long time).

Functional failures of some discrete elements in the system, such as these, may not change the system's functionality, provided **redundancy** is built into the system that leads to the system **reliability**. In general, redundancy increases reliability of the system, resulting in a tolerance of failures that may occur in the system. That means that the system as a whole performs its function, even if elements fail.

In 1952 John von Neumann presented a theory on building reliable systems from unreliable elements. His recipe for the design of such systems is outlined earlier in this paper, together with Fialkowski's claim dating from 1978 that evolution simply applied the solution discovered by von Neumann's to create heat-resistant brain of Homo; the brain worked reliably as a whole, despite its elements (neurons) losing their functionality as a result of overheating.

In functional terms, both our brains, and those of the earliest Homo are in one of the three zones, which like traffic lights, we shell term: "green", "amber" and "red". With negligible overheating of neurons as in the course of everyday life, the brain remains in the green zone. With an increase of overheating, the functional status of the brain shifts to the amber zone. In the amber zone some neurons lose their functionality, however, owing to the redundancy that every brain displays to a certain degree, the brain, as a whole does not change its functionality. Overall, it works as if noting had happened – one hundred percent effectiveness. With a further increase in overheating, the functional status enters the red zone, were brain redundancy is insufficient to mask the effects of the dysfunction of too many neurons: the brain begins to work incorrectly (losing its ability to control the body –also known as heat stroke!). Continued heat stress in the red zone ultimately leads to the black zone: the lethal zone.

Appling von Neumann's principle, evolution **increased the amber zone at the cost of the red zone**. In other words, an individual with larger and more interconnected brain could hunt successfully, whereas, when overheated to the same extent, those with less redundant brain lost body control; they suffered a heat stroke and thus had to abandon the hunt.

The more successful the hunt, the greater amount of meat: this increases the chances of passing the hunter's genes on to the next generation. In that manner, the adaptation for a larger and more interconnected brain had progressed. The average volume of the brain in the linage of successful hunters grew exponentially.

It should be noted, however, that using the larger brain infrastructure for more advanced thinking followed the growth of the brain – albeit with a most substantial delay. The limited extent of advanced mental capabilities is evident in the lack of any substantial progress in tool production until very late *Homo erectus* (Wynn, 1988).

Exponential growth in brain volume lasting nearly two million years came to a conclusion in approximately 200 000 years ago. Just it was the emergence of speech that brought an end to exponential growth. With the advent of symbolic communication, other, more sophisticated patterns of co-operative hunting could be introduced. Repeating the quotation of Hayden (1981, after Bortz 1985): "Groups will hunt as individuals when they can and communally when they have to". Given the faculty of speech, the group (contrary to earlier hominids) could have hunted "communally when they [had] to" i.e. during difficult times when prey was scarce. It was precisely these difficult times that constituted the period of strongest selection (Foley, 1987). Thus, the selection pressure that had become too strong while individual hunting was relaxed via a more sophisticated hunting mode available to hunters after the emergence of speech. As a result, brain growth ceased to expand exponentially and the inflection point occurred on the exponential curve.

Lieberman (1991: 109, 250 respectively) gave an independent dating of the emergence of fully developed speech between 125000 and 250000 BP. It coincides with the inflection point. It confirms a prior prediction (Fialkowski **1990**: 188) derived from the heat stress hypothesis.

Speech was the milestone in hominid evolution. Its emergence concluded the exponential growth of brain volume in hominids and generated new social selection pressures in addition to enhancing those that already existed (Machiavellian Intelligence 1988; Dunbar 1993) that were followed by new adaptations and ultimately culture.

**Lowering of the larynx position.**

Speech could not have emerged without the lower larynx position that is specific to humans. Lieberman (1991, p.74) identified the lower larynx position as a step towards speech evolution. He wrote: "The lower larynx position probably evolved to facilitate mouth breathing, which is an advantage for aerobic activities".

Fialkowski, in his hypothesis, claimed that the **"aerobic activities" were just persistence hunting** forcing the larynx to the new position. Apart from the enlarged brain - **'the reliability thermal buffer'** - the **lowering of the larynx position** was the second crucial pre-adaptation to speech. The unique, bipedal persistence running rendered both of them. Both were results of extraordinary circumstances forcing bipedal ape to hunt on the savannah though it lacked fangs and claws.

C. J. Lumsden. & E. O. Wilson in their book *Promethean Fire: Reflections on the Origin of Mind* (1983, p. 160) wrote: "Some extraordinary set of circumstances – the prime movers of the origin of mind – must have existed to bring the early hominids **across the Rubicon and into the irreversible march of cultural evolution**."

C. J. Lumsden & E.O. Wilson also wrote (*ibid.*: 162):
"So the search for the elusive prime mover must turn in new directions. A physiologist, K. R. Fialkowski has nominated heat stress as a key factor. When early men began hunting on the African savannas (according to this hypothesis), they lacked adequate cooling mechanisms for the brain. An excessively high blood temperature damaged nerve cells and impaired mental ability. The species responded by evolving larger brains with less densely packed and hence more redundantly acting nerve cells."

**Homo was brought across the Rubicon by chance.**